\documentclass[twocolumn,showpacs,preprintnumbers]{revtex4}

\usepackage{graphicx}
\usepackage{dcolumn}
\usepackage{bm}

\begin{document}


\title{Observation of Fragile-to-Strong Dynamic Crossover in Protein Hydration Water}

\author{Sow-Hsin Chen,$^{1}$\footnote{Author to whom correspondence should be
addressed. Electronic mail: sowhsin@mit.edu} Li Liu,$^{1}$
Emiliano Fratini,$^{2}$ Piero Baglioni,$^2$ Antonio Faraone,$^{3}$
and Eugene Mamontov$^{3}$ }
\address{$^{1}$Department of Nuclear Science and Engineering, Massachusetts Institute
of Technology, Cambridge MA 02139 USA $^{2}$Department of
Chemistry and CSGI, University of Florence,\\ via della Lastruccia
3, 50019 Florence, Italy $^{3}$Department of Material Science and
Engineering, University of Maryland, College Park, MD 20742 USA
and NIST Center for Neutron Research, Gaithersburg, MD 20899-8562
USA}


\begin{abstract}
At low temperatures proteins exist in a glassy state, a state
which has no conformational flexibility and shows no biological
functions. In a hydrated protein, at and above 220 K, this
flexibility is restored and the protein is able to sample more
conformational sub-states, thus becomes biologically functional.
This `dynamical' transition of protein is believed to be triggered
by its strong coupling with the hydration water, which also shows
a similar dynamic transition. Here we demonstrate experimentally
that this sudden switch in dynamic behavior of the hydration water
on lysozyme occurs precisely at 220 K and can be described as a
Fragile-to-Strong dynamic crossover (FSC).  At FSC, the structure
of hydration water makes a transition from predominantly
high-density (more fluid state) to low-density (less fluid state)
forms derived from existence of the second critical point at an
elevated pressure.
\end{abstract}


\maketitle

Without water, a biological system would not function. Dehydrated
enzymes are not active, but a single layer of water surrounding
them restores their activity. It has been shown that the enzymatic
activity of proteins depends crucially on the presence of at least
a minimum amount of solvent water~\cite{Gregory95,Teeter91}. It is
believed that about 0.3 $g$ of water per $g$ of protein is
sufficient to cover most of the protein surface with one single
layer of water molecules and to fully activate the protein
functionality. Thus, biological functions~\cite{Rupley80}, such as
enzyme catalysis, can only be understood with a precise knowledge
of the behavior of this single layer of water and how that water
affects conformation and dynamics of the protein. The knowledge of
the structure and dynamics of water molecules in the so-called
hydration layer surrounding proteins is, therefore, of utmost
relevance to the understanding of the protein functionality. It is
well documented that at low temperatures proteins exist in a
glassy state~\cite{Iben89,Angel95}, which is a solid-like
structure without conformational flexibility. As the temperature
is increased, the atomic motional amplitude increases linearly
initially, as in a harmonic solid. In hydrated proteins, at
approximately 220 K, the rate of the amplitude increase suddenly
becomes enhanced, signalling the onset of additional anharmonic
and liquid-like
motion~\cite{Parak84,Doster2,Doster90,Rasmussen92}. This
`dynamical' transition of proteins is believed to be triggered by
their strong coupling with the hydration water through the
hydrogen bonding. The reasoning is derived from the finding that
the protein hydration water shows some kind of dynamic transition
at the similar temperature~\cite{Paciaroni99,Caliskan02}. Here we
demonstrate, through a high-resolution quasi-elastic neutron
scattering (QENS) experiment, that this dynamic transition of
hydration water on lysozyme protein is in fact the
Fragile-to-Strong dynamic crossover (FSC) at 220 K, similar to
that recently observed in confined water in cylindrical nanopores
of silica materials~\cite{Faraone04,LiuPRL}. Computer simulations
on both bulk water~\cite{XuPNAS} and protein hydration water
around lysozyme~\cite{KumarSubmit} have led to the interpretation
of the FSC as arising from crossing the locus of maximum in the
correlation length (``Widom line") which emanates from a critical
point into the one-phase region; if this interpretation is
correct, then our experiments provide evidence supporting the
existence of a liquid-liquid critical point in protein hydration
water, which previously has been proposed only for bulk
water~\cite{Gregory}.

Water molecules in a protein solution may be classified into three
categories: (i) the bound internal water, (ii) the surface water,
i.e. the water molecules that interact with the protein surface
strongly, and (iii) the bulk water. The bound internal water
molecules, which occupy internal cavities and deep clefts, are
extensively involved in the protein-solvent H-bonding, and play a
structural role in the folded protein itself. The surface water,
or usually called the hydration water, is the first layer of water
that interacts with the solvent-exposed protein atoms of different
chemical character, feels the topology and roughness of the
protein surface, and exhibits the slow dynamics. Finally, water,
which is not in direct contact with the protein surface but
continuously exchanges with the surface water, has properties
approaching that of bulk water. In this letter, we deal with
dynamics of the hydration water in a powder of the globular
protein lysozyme. This hydration water is believed to have an
important role in controlling the bio-functionality of the
protein.

The biochemical activity of proteins also depends on the level of
hydration. In lysozyme, enzymatic activity remains very low up to
a hydration level ($h$) of $\approx$ 0.2 ($h$ is measured in $g$
of water per $g$ of dry protein) and then increases sharply with
an increase in $h$ from 0.2 to 0.5~\cite{Rupley91}. Various
experiments~\cite{Roh05} and computer simulations~\cite{Tarek02}
have demonstrated the strong influence of the hydration level on
protein dynamics.

It has been found that many proteins exhibit a sharp slowing down
of their functions (kinetic of biochemical reactions) at a
temperature somewhere within the interval of $T$ between 200 K and
250 K. An analysis of the mean-squared atomic displacement,
$\langle x^2 \rangle$, by using Mössbauer~\cite{Parak81},
X-ray~\cite{Rasmussen92}, and neutron scattering~\cite{Ferrand93}
spectroscopy, in hydrated proteins shows sharp changes around a
certain sample temperature range: $\langle x^2 \rangle$ varies
approximately linearly as a function of $T$ at low $T$ and then
increases sharply above $T$ between 200~K and 250~K. The sharp
rise in $\langle x^2 \rangle$ was attributed to a certain dynamic
transition in biopolymers at this temperature range. The
coincidence of the characteristic temperatures, below which the
biochemical activities slow down, and the on-set of the dynamic
transition, suggests a direct relation between these two
phenomena. It has also been demonstrated that the dynamic
transition can be suppressed in dry biopolymers, or in biopolymers
dissolved in trehalose~\cite{Cordone99}. It can also be shifted to
higher temperatures, e.g. between 270~K and 280~K, for proteins
dissolved in glycerol~\cite{Tsai00}. Thus the solvent plays a
crucial role in the dynamic transition in biopolymers. This
observation led to a suggestion by many authors that proteins are
`slaves' to the solvent~\cite{Lich99}. Despite many experimental
studies, the nature of the dynamic transition in proteins remains
unclear. Many authors interpret the dynamic transition as a kind
of glass transition in a
biopolymer~\cite{Doster2,Doster90,Vitkup00}. Our experiments
described below demonstrate that the origin of the characteristic
temperature controlling both the activity of the protein and the
transition in the behavior of $\langle x^2 \rangle$ is the FSC
phenomenon in the hydration water, which shares the same crossover
temperature with the protein.

Using high-resolution QENS method and the Relaxing-Cage Model
(RCM, described in Methods section)~\cite{Chen99} for the
analysis, we determine the temperature dependence of the average
translational relaxation time, $\left\langle\tau_T\right\rangle$,
for the hydration water. The dynamic crossover temperature of
hydration water is defined as follows. At high temperatures,
$\left\langle\tau_T\right\rangle$ follows a super-Arrhenius
behavior (called a `fragile' behavior~\cite{Angell91}) describable
approximately by a Vogel-Fulcher-Tammann (VFT) law:
$\left\langle\tau_T\right\rangle=\tau_1$exp$\left[DT_0/(T -
T_0)\right]$, where $D$ is a constant providing the measure of
fragility and $T_0$, the ideal glass transition temperature at
which the relaxation time appears to diverge. In reality, however,
this divergence is avoided by the system. Instead, an Arrhenius
behavior (called a `strong' behavior~\cite{Angell91}) sets in
below the crossover temperature $T_L$ where the functional
dependence of the relaxation time switches to a law:
$\left\langle\tau_T\right\rangle=\tau_1$exp$\left[E_A/RT\right]$.
In this equation, $E_A$ is the activation energy for the
relaxation process and $k_B$, the Boltzmann constant. The
crossover temperature $T_L$ is defined by the intersection of
these two laws, which gives $1/T_L = 1/T_0 - (Dk_B)/E_A$. In the
case of hydration water in lysozyme, we found $T_L$~=~220 K, which
agrees well with the characteristic transition temperature in
protein observed before~\cite{Rasmussen92}. Since the average
relaxation time $\left\langle\tau_T\right\rangle$ is a measure of
the mobility of a typical hydration water molecule, this result
implies that the sudden change in the trend of mobility of water
molecules at the crossover temperature triggers the so-called
glass transition of protein
molecules~\cite{Doster2,Doster90,Tarek02,Vitkup00}.

\section*{Results}

\begin{figure}[tbp]
\begin{center}
\includegraphics[width=8.5 cm]{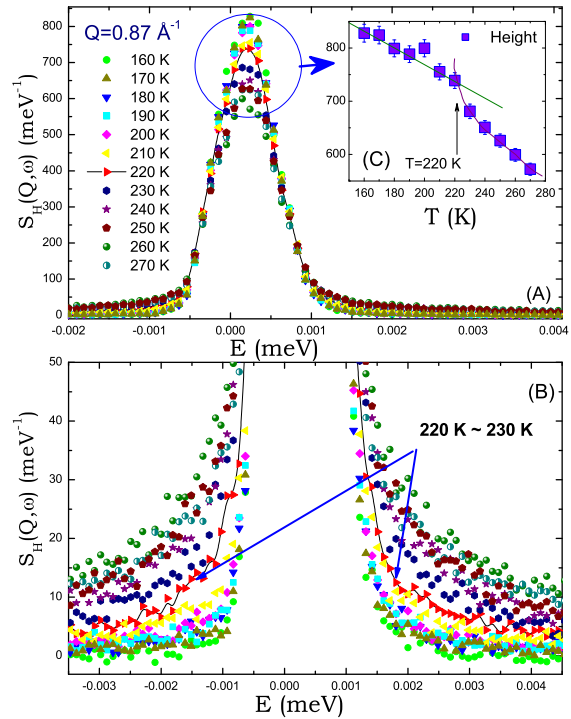}
\end{center}
\caption{{{{{{{{{{{{\protect\small Measured neutron spectra.
Panels A and B show normalized QENS spectra at $Q =
0.87$~\AA$^{-1}$, displaying the height of the peak (A) and the
wing of the peak (B), respectively, at a series of temperatures.
The inset C shows a plot of the peak heights versus temperature.
Arrow signs in B are intended to highlight the wing of the spectra
at the crossover temperature. }}}}}}}}}}}} \label{fig1}
\end{figure}

We show in Fig.~\ref{fig1}, as an example, a complete set
(temperature series) of QENS area-normalized spectra taken at $Q =
0.87$~\AA$^{-1}$ at ambient pressure. The broadening of the
quasi-elastic peaks at the wing becomes more and more noticeable
as temperature increases. And at the same time, the peak height
decreases accordingly because the area is normalized to unity. In
the inset C, we plot the peak height as a function of temperature.
It is noticeable that the rate of increase as a function of
temperature is different across the temperature 220~K. From
Fig.~\ref{fig1}B, we may notice, from wings of these spectral
lines, that two groups of curves, 270~K~$\geq$~T~$\geq$~230~K and
210~K~$\geq$~T~$\geq$~160~K, are separated by the curve at a
temperature 220 K. This visual information obtained from the
spectra before data analysis reinforces the result of the detail
line shape analysis to be shown later in Fig.~\ref{fig3}, that
there is an abrupt dynamical transition at $T_L$~=~220~K.

\begin{figure}[tbp]
\begin{center}
\includegraphics[width=8.5 cm]{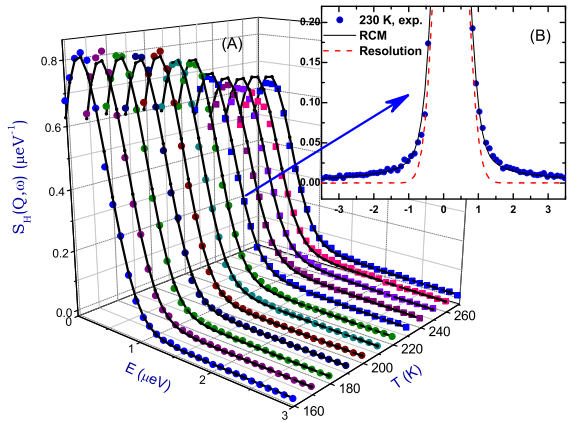}
\end{center}
\caption{{{{{{{{{{{{\protect\small Neutron spectra and their RCM
analyses. Panel A displays measured QENS spectra (solid symbols)
and their RCM analysis results (solid lines) at $Q =
0.87$~\AA$^{-1}$ and at a series of temperatures. Panel B singles
out one particular spectrum at T~=~230~K and contrasts it with the
resolution function of the instrument for this Q value (dash
line). }}}}}}}}}}}} \label{fig2}
\end{figure}

Fig.~\ref{fig2} shows the result of RCM analyses of the spectra
taken at $Q = 0.87$~\AA$^{-1}$ for a series of temperatures
ranging from 270~K to 160~K (panel A), and in particular at
T~=~230~K (panel B). In this figure, we display the instrument
resolution function purposely for comparison with the measured
spectrum. RCM, as one can see, reproduces the experimental
spectral line shapes of hydration water quite well. The broadening
of the experimental data over the resolution function leaves
enough dynamic information to be extracted by RCM. This means that
it requires a high-resolution backscattering instrument with an
energy resolution of 0.8~$\mu$eV to adequately study the FSC
phenomena in hydration water.

\begin{figure}[tbp]
\begin{center}
\includegraphics[width=8.5 cm]{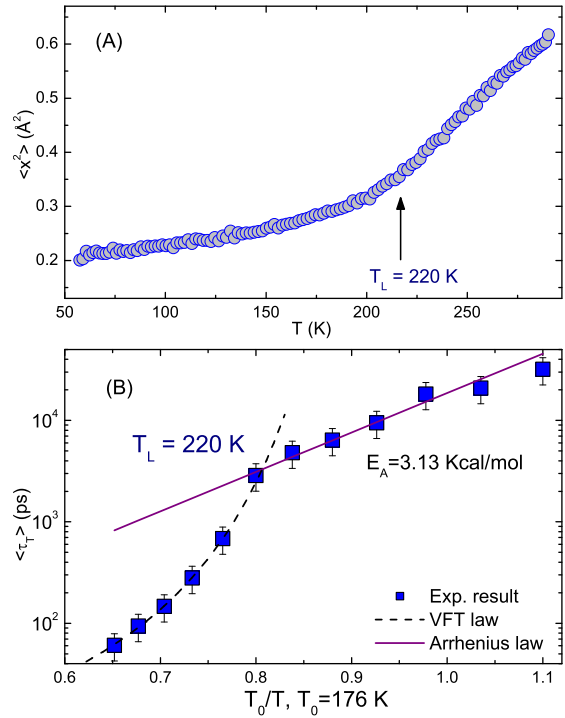}
\end{center}
\caption{{{{{{{{{{{{\protect\small Evidence for the dynamic
transition. Panel A shows the temperature dependence of the
mean-squared atomic displacement of the hydrogen atom at 2
nanosecond time scale measured by an elastic scan with resolution
of 0.8 $\mu$eV. Panel B shows temperature dependence of the
average translational relaxation times plotted in log$(\langle
\tau_T \rangle)$ versus $T_0/T$, where $T_0$ is the ideal glass
transition temperature. In this figure, there is a clear and
abrupt transition from a VFT law at high temperatures to an
Arrhenius law at low temperatures, with the fitted crossover
temperature $T_L$ = 220 K and the activation energy $E_A$ =
3.13~kcal/mol extracted from the Arrehenius part indicated in the
figure. }}}}}}}}}}}} \label{fig3}
\end{figure}

In Fig.~\ref{fig3}, we first present (in panel A) the mean-squared
atomic displacement $\langle x^2 \rangle$ of the hydrogen atoms
(calculated from the translational Debye-Waller factor,
$S_H(Q,\omega =0)=$~exp$[-Q^2 \langle x^2 \rangle ]$) versus $T$
to indicate that there is a hint of a dynamic transition at a
temperature between 200~K and 220~K. In panel B, we then present
in an Arrhenius plot the temperature dependence of the average
translational relaxation time, $\langle \tau_T \rangle$, for the
hydrogen atom in a water molecule calculated by Equs. (1-2) in the
Methods section. The contribution from hydrogen atoms in the
protein has been subtracted out during the signal processing. It
is seen that, in the temperature range 270-230~K, $\langle \tau_T
\rangle$ obeys VFT law, a signature of fragile liquid, quite
closely. But at T~=~220~K it suddenly switches to an Arrhenius
law, a signature of a strong liquid. So we have a clear evidence
of FSC in a cusp form. The $T_0$ for the fragile liquid turns out
to be 176~K, and the activation energy for the strong liquid,
$E_A=$~3.13~kcal/mol.

\section*{Discussion and Summary}

Recently, E. Mamontov observed a similar dynamic crossover in the
surface water on cerium oxide powder sample~\cite{Mamontov05}. The
surface of cerium oxide is hydroxylated. The coverage of water is
about 2 layers and the crossover temperature is said to be at 215
K. The observed slow dynamics is attributed to the effect on the
translational mobility of the water molecules in contact with the
surface hydroxyl groups. Thus, our observation of the FSC in
hydration water of protein may be a universal phenomenon for
surface water.

It should be noted that the FSC in confined supercooled water is
attributed to the crossing of the so-called Widom line in the
Pressure-Temperature (phase) plane in a recent MD simulation work
on bulk water~\cite{XuPNAS} and protein hydration water at ambient
pressure~\cite{KumarSubmit}. The Widom line is originated from the
existence of the second critical point of water and is the
extension of the liquid-liquid coexistence line into the one phase
region. Therefore, our observation of the FSC at ambient pressure
implies that there may be a liquid-liquid phase transition line in
the protein hydration water at elevated pressures. This dynamic
crossover, when crossing the Widom line, causes the layer of the
water surrounding a protein to change from the `more fluid'
high-density liquid form (which induces the protein to adopt more
flexible conformational sub-states) to the `less fluid'
low-density liquid structure (which induces the protein to adopt
more rigid conformational sub-states).

In summary, an investigation of the average translational
relaxation time, or the alpha-relaxation time, of protein
hydration water as a function of temperature reveals a hitherto
un-noticed Fragile-to-Strong dynamic crossover at 220~K, close to
the universal dynamic transition temperature documented for
proteins in literature. This fact implies that the sudden
transition of the water mobility on the surface of a protein at
the FSC triggers the so-called glass transition, which is known to
have a profound consequence on biological function of the protein
itself.

\section*{Methods}
\subsection*{Sample preparation}
Hen egg white lysozyme used in this experiment was obtained from
Fluka (L7651, three times crystallized, dialysed and lyophilized)
and used without further purification. The sample was dried under
vacuum in the presence of P$_2$O$_5$ to remove any water left. The
dried protein powder was then hydrated isopiestically at
5~$^{\circ}$C by exposing it to water vapor in a closed chamber
until $h$~=~0.3 is reached (i.e. 0.3 $g$ H$_2$O  per $g$ dry
lysozyme). The hydration level was determined by
thermo-gravimetric analysis and also confirmed by directly
measuring the weight of absorbed water. This hydration level was
chosen to have almost a monolayer of water covering the protein
surface~\cite{Careri98}. A second sample was then prepared using
D$_2$O in order to subtract out the incoherent signal from the
protein hydrogen atoms. Both hydrated samples had the same water
or heavy water/dry protein molar ratio. Differential scanning
calorimetry (DSC) analysis was performed in order to detect the
absence of any feature that could be associated with the presence
of bulk-like water.

\subsection*{Neutron experiments}
High-resolution incoherent QENS spectroscopy method is used to
determine the temperature dependence of the average translational
relaxation time for the hydration water. In this experiment, we
measured both an H$_2$O hydrated sample and a D$_2$O hydrated
sample, and take the difference to obtain the signal contributed
solely from hydration water. Because neutrons are predominantly
scattered by an incoherent process from the hydrogen atoms in
water (rather than by the coherent scattering process from the
oxygen atoms), high-resolution QENS technique is an appropriate
tool for the study of diffusional process of water molecules.
Using the High-Flux Backscattering Spectrometer (HFBS) in NIST
Center for Neutron Research (NCNR), we are able to measure the
average translational relaxation time ($\alpha$-relaxation time)
from 60 $ps$ to 20 $ns$ over the temperature range of 270~K to
180~K, spanning both below and above the FS crossover temperature.
For the chosen experimental setup, the spectrometer has an energy
resolution of $\pm$ 0.8 $\mu$eV and a dynamic range of
$\pm$~11~$\mu$eV~\cite{Meyer03}, in order to be able to extract
the broad range of relaxation times covering both the fragile and
the strong regimes of the relaxation times from measured spectra.

\subsection*{Data Analysis}

\begin{figure}[tbp]
\begin{center}
\includegraphics[width=8.5 cm]{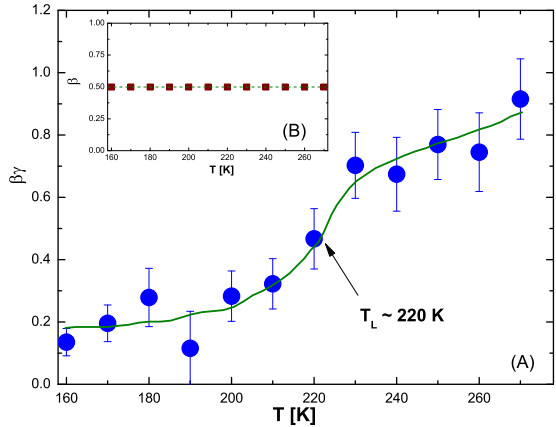}
\end{center}
\caption{{{{{{{{{{{{\protect\small Temperature dependence of the
exponents $\beta$ and $\beta \gamma$ giving respectively the power
laws of$t$ and $Q$-dependence of ISF.  }}}}}}}}}}}} \label{fig4}
\end{figure}

QENS experiments essentially provide us with the Fourier transform
of the Intermediate Scattering Function (ISF) of the hydrogen
atoms, $F_H (Q,t)$, of water molecules in the hydration layer. MD
simulations have shown that the ISF of both bulk~\cite{Gallo96}
and confined~\cite{Gallo00} supercooled water can be accurately
described as a two-step relaxation: a short-time Gaussian-like
(in-cage vibrational) relaxation followed by a plateau and then a
long-time (time longer than 1.0 $ps$) stretched exponential
relaxation of the cage. The RCM~\cite{Chen99}, which we use for
data analysis, models closely this two-step relaxation and has
been tested extensively against bulk and confined supercooled
water through MD and experimental
data~\cite{Gallo96,Gallo00,Faraone04}. By considering only the
spectra with wave vector transfer $Q < 1.1$~\AA$^{-1}$, we can
safely neglect the contribution from the rotational motion of
water molecule in ISF~\cite{Chen99}. The RCM describes the
translational dynamics of water at supercooled temperature in
terms of the product of two functions:

\begin{eqnarray}
F_H\left(Q,t\right)&\approx& F_T\left(Q,t\right)
=F^S\left(Q,t\right)exp\left[-\left(t/\tau_T(Q)\right)^\beta\right], \nonumber \\
\tau_T\left(Q\right)&=&\tau_0\left(0.5Q\right)^{-\gamma}
\label{decoupling}
\end{eqnarray}

\noindent where the first factor, $F^S\left(Q,t\right)$,
represents the short-time vibrational dynamics of the water
molecule in the cage. This function is fairly insensitive to
temperature variation, and thus can be calculated from MD
simulation. The second factor, the $\alpha$-relaxation term,
contains the stretch exponent $\beta$, and the $Q$-dependent
translational relaxation time $\tau_T (Q)$, which is a strong
function of temperature. The latter quantity is further specified
by two phenomenological parameters $\tau_0$ and $\gamma$, the
exponent controlling the power-law $Q$-dependence of
$\tau_T\left(Q\right)$. The average translational relaxation time,
which is a $Q$-independent quantity, we use in this paper is
defined as:

\begin{equation}
\left\langle\tau_T\right\rangle=\tau_0\Gamma\left(1/\beta\right)/\beta,
\label{Equa2}
\end{equation}

\noindent where $\Gamma$ is the gamma function. The temperature
dependence of the three phenomenological parameters, $\tau_0$,
$\beta$, and $\gamma$, are obtained by analyzing simultaneously a
group of nine quasi-elastic peaks at different $Q$ values. Then
the average translational relaxation time, $\langle \tau_T
\rangle$, is calculated according to Equ. (2) using $\tau_0$ and
$\beta$. As can be seen from Equs. (1) and (2), the product $\beta
\gamma$ is an exponent expressing the $Q$-dependence of the ISF.
Fig.~\ref{fig4} gives the temperature dependence of the exponent
$\beta \gamma$, which indicates a precipitous drop at the
crossover temperature 220~K from a high temperature value of about
1 to a low temperature value of about 0.2, while $\beta$ maintains
a value of 0.5 all the way through~\cite{Faraone04}. Note that the
values of exponents $\beta$ and $\beta \gamma$ for free diffusion
are 1 and 2, respectively, and the lower values of these two
exponents signify an anomalous and more restrictive mobility of
water compared to that of the free diffusion. The Fig.~\ref{fig4}
shows that while the mobility of hydration water molecule at $ps$
to $ns$ timescale deviates significantly from free diffusion at
all temperatures measured, as temperature goes below the crossover
temperature $T_L$, the mobility becomes drastically reduced.

\begin{center}
{\bf Acknowledgement }
\end{center}

We thank Dr. H. Eugene Stanley for a critical review of the
manuscript. The research at MIT is supported by DOE Grants
DE-FG02-90ER45429 and 2113-MIT-DOE-591. This work utilized
facilities supported in part by the National Science Foundation
under Agreement No. DMR-0086210. We benefited from affiliation
with EU-Marie-Curie Research and Training Network on Arrested
Matter.

\end{document}